\documentclass[preprint,amsmath,amssymb]{revtex4}
\usepackage{graphicx}% Include figure files
\usepackage{dcolumn}% Align table columns on decimal point
\usepackage{bm}% bold math
\usepackage{natbib}

\begin{document}
\title{Lagrangian one-particle velocity statistics in a turbulent flow}
\author{Jacob Berg}
\email{jacob.berg.joergensen@risoe.dk}

\affiliation{Ris{\o} National Laboratory, 4000 Roskilde, Denmark}

\date{\today}
\begin{abstract}
We present Lagrangian one-particle statistics from the Ris{\o} PTV
experiment of a turbulent flow. We estimate the Lagrangian
Kolmogorov constant $C_0$ and find that it is affected by the
large scale inhomogeneities of the flow. The pdf of temporal
velocity increments are highly non-Gaussian for small times which
we interpret as a consequence of intermittency. Using Extended
Self-Similarity we manage to quantify the intermittency and find
that the deviations from Kolmogorov 1941 similarity scaling is
larger in the Lagrangian framework than in the Eulerian. Through
the multifractal model we calculate the multifractal dimension
spectrum.

\end{abstract}

\maketitle
\section{Introduction}
In the present contribution we present experimental results on
Lagrangian one-particle statistics from an experiment with the
Ris{\o} Particle Tracking Velocimetry (PTV) setup. We focus on
small-scale statistics in a turbulent flow: the statistic is
analyzed with Extended Self-Similarity (ESS)~\cite{benzi:1993} and
the results are presented in the spirit of the multifractal model
of turbulence~\citep{frisch}. The use of ESS is discussed together
with the multifractal model in a finite Reynolds number flow like
the present.

\begin{figure}[h]
\includegraphics[width=0.8\columnwidth]{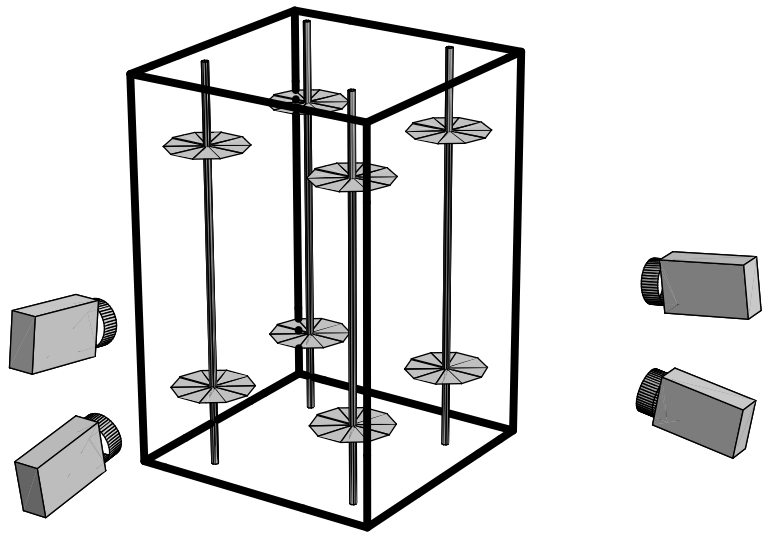}\vspace{-2cm}
\caption{{\footnotesize Experimental setup}} \label{fig:exp}
\end{figure}

We have performed a Particle Tracking Velocimetry (PTV) experiment
in an intermediate Reynolds number turbulent flow. The flow has
earlier been reported in \cite{berg:2006,luthi:2006b,berg:2006b}
although we use data from a recording with a slightly lower
Reynolds number. PTV is an experimental method suitable for
obtaining Lagrangian statistics in turbulent flows
\citep{ott:2000,laporta:2001,mordant:2001,luthi:2005,cornell,xu:2006}:
Lagrangian trajectories of fluid particles in water are obtained
by tracking neutrally buoyant particles in space and time. The
flow is generated by eight rotating propellers, which change their
rotational direction in fixed intervals in order to suppress a
mean flow, placed in the corners of a tank with dimensions
$32\times32\times50\mathrm{cm}^3$ (see Fig~\ref{fig:exp}).
The data acquisition system consists of four commercial CCD
cameras with a maximum frame rate of $50\mathrm{Hz}$ at $1000
\times 1000$ pixels. The measuring volume covers roughly
$(12\mathrm{cm})^3$. We use polystyrene particles with size
$\sim400\mathrm{\mu m}$ and density very close to one. We follow
$\mathcal{O}(1000)$ particles at each time step with a position
accuracy of $0.05$ pixels corresponding to less than $10\mathrm{
\mu m}.$

The Stokes number, $\tau_I / \tau_{\eta}$ ($\tau_I$ denotes the
inertial relaxation time for the particle to the flow while
$\tau_{\eta}$ is the Kolmogorov time) is much less than one and
the particles can therefore be treated as passive tracers in the
flow. The particles are illuminated by a $250\mathrm{W}$ flash
lamp.

The mathematical algorithms for translating two dimensional image
coordinates from the four camera chips into a full set of three
dimensional trajectories in time involve several crucial steps:
fitting gaussian profiles to the 2d images, stereo matching (line
of sight crossings) with a two media (water-air) optical model and
construction of 3d trajectories in time by using the kinematic
principle of minimum change in acceleration
\citep{willneff:2003,ouellette:2006}.

The flow characteristics are presented in Table~\ref{table:flow}.
The mean flow is axisymmetric with a significant vertical
straining on the largest scales and we did not find any
significant differences from the flow reported in
\citep{berg:2006,luthi:2006b}, where properties of the mean flow
can be found.

Here we look at a sub-volume of the full measuring volume. Only
particles which we can observe within a ball of radius
$50\mathrm{mm}$ is considered and the turbulence characteristics
given in Table~\ref{table:flow} are thus only determined from
particles inside this ball. The ball is centered approximately in
the center of the tank where the velocity standard deviation
$\sigma_u$ has a global minimum. Inside the ball the particles are
uniformly distributed. With $\tau_{\eta}=0.09s$ and a recording
frequency at $50Hz$ the temporal resolution is $\sim4\mathrm{
frames}/\tau_{\eta}$.

\begin{table}[t]
\begin{tabular}{|c|c|c|c|c|c|c|}
  \hline
  $\eta$ & $L$ & $\tau_{\eta}$ & $T_E$ &$\varepsilon$ & $\sigma_u$ & $Re_{\lambda}$ \\
  \hline\hline
  $0.30\mathrm{mm}$ & $53.80\mathrm{mm}$ & $0.09\mathrm{s}$ & $2.83\mathrm{s}$& $128\mathrm{mm^2/s^3}$ & $19.02\mathrm{mm/s}$ & 124 \\
  \hline
\end{tabular}
\caption{{\footnotesize Turbulence characteristics: $\varepsilon$
is the mean kinetic energy dissipation,
$\eta\equiv(\nu^3/\varepsilon)^{1/4}$ is the Kolmogorov length
scale with the kinematic viscosity $\nu=1$ of water.
$\tau_{\eta}\equiv (\nu/\varepsilon)^{1/2}$ is the Kolmogorov time
scale and
$\sigma_u^2=\frac{1}{3}(\sigma_{u_x}^2+\sigma_{u_y}^2+\sigma_{u_z}^2)$
is the standard deviation of velocity. The integral length scale
is defined as $L=\sigma^3/\varepsilon$ while $T_E$ is the eddy
turnover time $T_E=L/\sigma_u$. The Reynolds number is defined as
$Re_{\lambda}=\sqrt{15} (L/\eta)^{2/3}.$}} \label{table:flow}
\end{table}

The database is the largest we have compiled and it consists of
$\sim10^6 $ individual trajectories with an average length of
$\sim 8\tau_{\eta}$, a standard deviation of $\sim 13\tau_{\eta}$
and the longest tracks we find are $\sim 150\tau_{\eta}$. The
number of tracks was an important requirement since the
calculation of high order moments is considered important.

Throughout the paper we will denote the \textit{Lagrangian}
velocity along a particle trajectory for $\mathbf{v}(t)$ and the
\textit{Eulerian} velocity in a fixed frame of reference for
$\mathbf{u}(\mathbf{x},t)$.

\section{Finite Volume measurements}
A nice property of the Eulerian velocity statistic is that it is
stationary in time in the present experiment. This is not the case
for the Lagrangian statistics.
\begin{figure}[h]
\begin{center}
\includegraphics[width=\columnwidth]{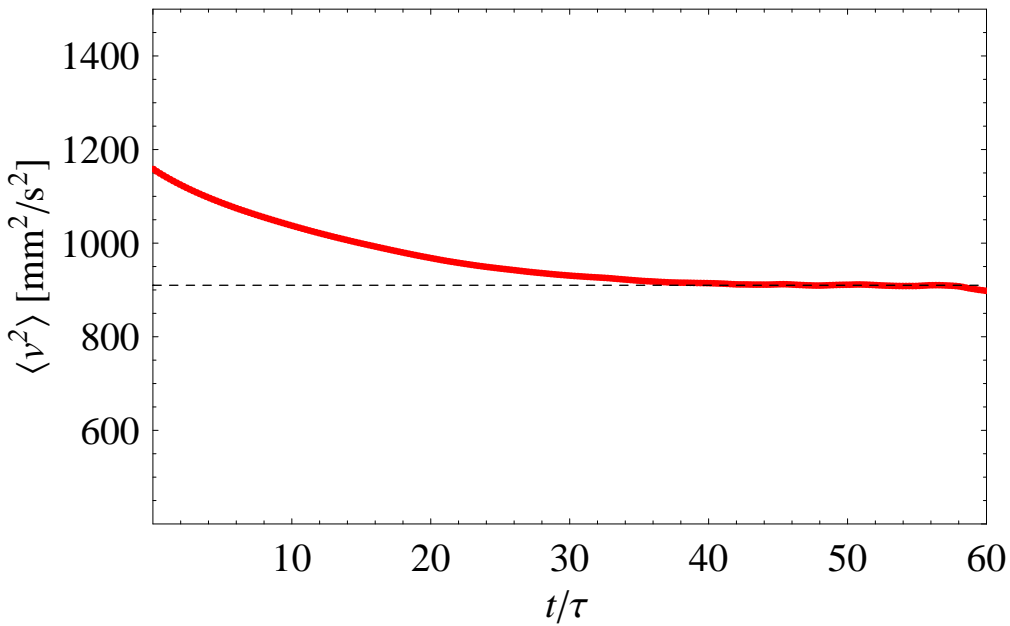}
\caption{{\footnotesize $\langle v^2(t+t_0) \rangle$. The average
is taken over all particles which were observed inside $B$ at both
time $t_0$ and $t_0+t$.}} \label{fig:v2}
\end{center}
\end{figure}
The non-stationarity is showed in Figure~\ref{fig:v2} where
$\langle v^2 \rangle$ is observed to decrease over time. This
reflects the finite measuring volume and the non-uniform forcing
in space in our experiment: the particles only gain kinetic energy
close to the propellers. During their subsequent motion the
particles loose kinetic energy until they again come close to the
propellers which are constantly spinning. Looking at a finite
measuring volume away from the propellers, there will therefore be
a flux of kinetic energy into the volume. Inside the volume the
kinetic energy is dissipated and hence we have at the entry of the
volume
\begin{equation}
\frac{1}{2} \frac{d}{dt} \langle v^2 \rangle = -\varepsilon,
\label{eq:v2}
\end{equation}
where we recognize the mean kinetic energy dissipation
$\varepsilon$. From Figure~\ref{fig:v2} we find
$\varepsilon=124\mathrm{mm^2/s^2}$. This number is close to the
number obtained from the second order Eulerian Structure Function
$\varepsilon=132\mathrm{mm^2/s^2}$. We take the difference as the
uncertainty in estimating $\varepsilon$. Eqn.~\ref{eq:v2} can also
be derived directly from the Navier-Stokes equation by assuming
global homogeneity.

\begin{figure}[h]
\begin{center}
\includegraphics[width=\columnwidth]{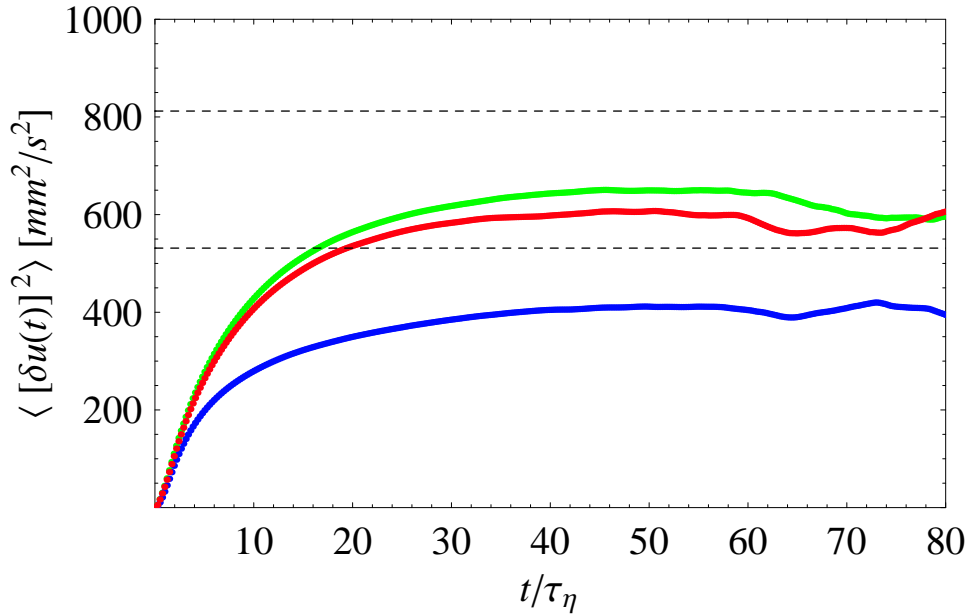}
\caption{{\footnotesize Second order Lagrangian structure function
for the three coordinates of $\mathbf{v}(t)$. $x:$ green (radial
component), $y:$ red (radial component) and $z:$ blue (vertical
component). The horizontal lines is the Eulerian velocity variance
$\langle u^2 \rangle$.}} \label{fig:s2}
\end{center}
\end{figure}

The Lagrangian second-order structure function is defined as
\begin{equation}
S_L^2(\tau)=\langle [v(t+\tau)-v(t)]^2 \rangle,
\end{equation}
where $v(t)$ is here the velocity component along a fluid
trajectory. Similar the Lagrangian co-variance function is defined
as
\begin{equation}
R_L(\tau)=\langle v(t) v(t+\tau) \rangle.
\end{equation}
The non-stationarity of $\langle v^2 \rangle$ means that
\begin{equation}
S^2_L(t) = \langle v^2(t) \rangle + \langle u^2\rangle -2 R_L(t) <
2 (\langle u^2 \rangle - R_L(t)), \label{eq:nonStationarity}
\end{equation}
where we have used that the Lagrangian velocity on the boundary of
the measuring volume $B$ equals the Eulerian velocity and
therefore $\langle v^2(t=0) \rangle = \langle u^2 \rangle$.
$S^2_L(t)$ is plotted in Figure~\ref{fig:s2} for all three
velocity components.
It is clear that for long times $S^2_L$ does not approach
$2\langle u^2 \rangle$ in agreement with
eqn.~\ref{eq:nonStationarity}.

A common interpretation of the finite volume influence on
Lagrangian statistics is that the particles we can observe for
long times are relatively slow ones or particles which are trapped
in high intensity vortices (see later). Here we emphasize the
equivalence with the energy argument of decaying turbulence
described above. Particles which can be observed for long times
are slow because it is long time ago they gained kinetic energy at
the forcing site.

In Direct numerical simulations (DNS) forcing occurs in
wave-number space on the lowest wave-numbers. We therefore have
$d\langle v^2 \rangle/dt=0$ and consequently Lagrangian
stationarity. Most physical flows encountered in nature will,
however, be Lagrangian non-stationary.

\section{Anisotropy and inertial range scaling}
The linear dependence of $Re_{\lambda}$ on $T_L/\tau_{\eta}$
implies that a very high Reynolds number is needed in order to
obtain a clear Lagrangian inertial range. \citet{yeung:2002}
concluded, based on extrapolations from Eulerian fields in DNS,
that at least $Re_{\lambda}\sim 600-700$ was needed. Experimental
flows at $Re_{\lambda}=1000$~\cite{mordant:2004b} and
$Re_{\lambda}= 815$~\cite{ouellette:2006b} do, however, not show a
very pronounced inertial range defined as a linear regime in the
second-order structure function $S_L^2$.

\begin{figure}[h]
\begin{center}
\includegraphics[width=\columnwidth]{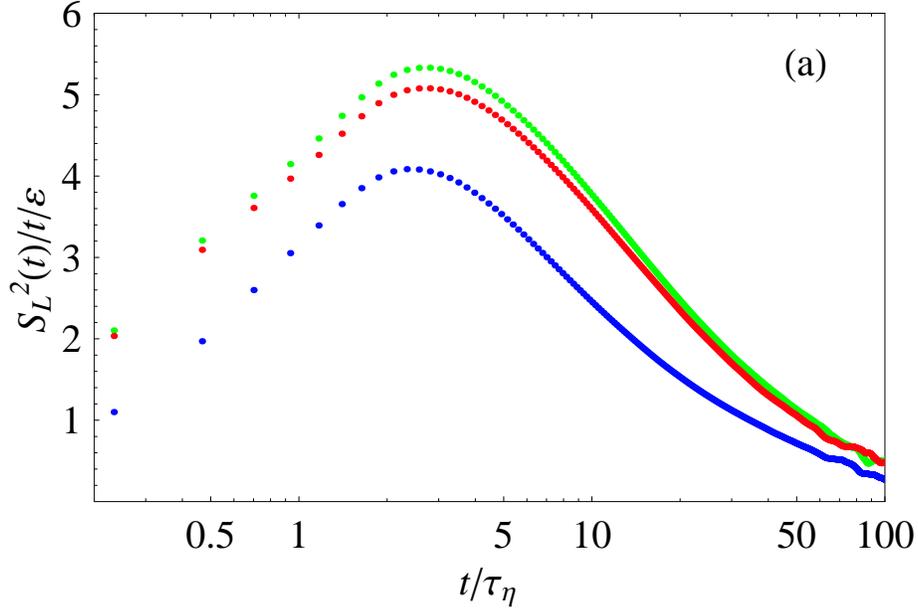}
\caption{{\footnotesize $C_0$ for the radial components (green and
red) and the axisymmetric component (blue).}} \label{fig:C0}
\end{center}
\end{figure}

In the inertial range $\tau_{\eta}<\tau<T_L$, K41 similarity
theory predicts
\begin{equation}
S^2_{L, ij}(\tau)=C_0 \varepsilon \tau \delta_{ij}, \label{eq:K41}
\end{equation}
where the Lagrangian Kolmogorov constant $C_0$ is supposed to be
universal for infinite Reynolds numbers~\citep{sawford:1991}.
$C_0$ plays a crucial role in stochastic
models~\citep{sawford:2001} and has lately been shown to reflect
anisotropy in the large-scale forcing \citep{ouellette:2006b}. In
Figure~\ref{fig:C0} we present results of $C_0$ for the three
coordinates of $\mathbf{v}(t)$.
According to eqn.~\ref{eq:K41}, $C_0$ should be determined from a
plateau in the inertial range. The parabolic form therefore
reflects the almost vanishing inertial range in our experiments.
The difference between radial and the axisymmetric component stems
from the large scale anisotropy. Since $C_0$ is maximum for times
around $2-4\tau$ and therefore mainly associated with small scales
the difference is a clear signature of small-scale anisotropy. The
values of $C_0$ are $5.34\pm0.16$, $5.08\pm0.15$ and $4.09\pm0.12$
for the three components $x$, $y$ and $z$ respectively.

It is interesting to see that the slight difference in the radial
forcing is surviving all the way down. The propellers forcing the
flow are attached to four rods placed in the corners of the tank.
The reason for the radial components being different is probably
small differences in the manual vertical placement of the
propellers on the rods. The lack of small-scale isotropy in the
current experiment should not be taken as a failure of
Kolmogorov's hypothesis of local isotropy. For that the Reynolds
number is not high enough. Other experiments at much higher
Reynolds number do, however, all indicate that the large scale
inhomogeneities are also present at smaller scales although with
smaller amplitude~\citep{shen:2000,shen:2002,ouellette:2006b}.

\begin{figure}[h]
\begin{center}
\includegraphics[width=\columnwidth]{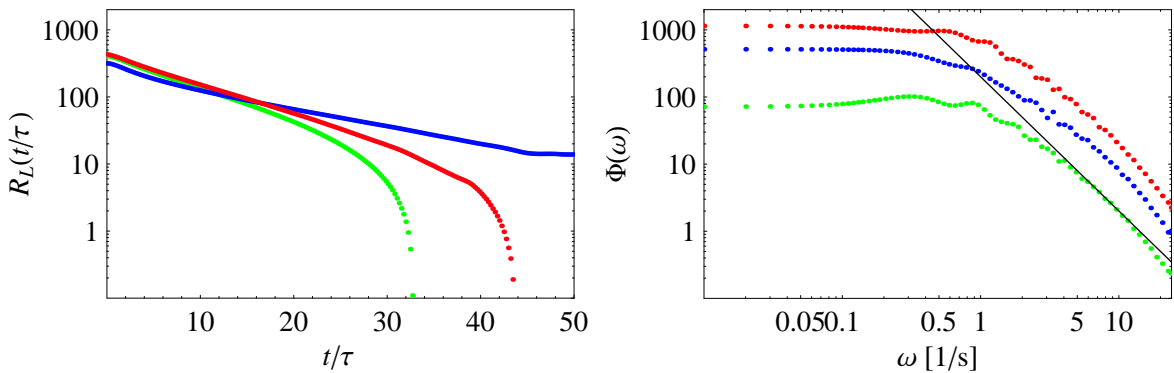}
\caption{{\footnotesize (a) $R_L(\tau)$. (b) $\Phi(\omega)$. The
straight line is the K41 prediction $\sim\omega^{-2}$. Color codes
as in Figure \ref{fig:C0}. The curves have been shifted vertically
for clarity.}} \label{fig:Ru}
\end{center}
\end{figure}

Alternatively one can calculate the lagrangian velocity spectrum
$\Phi(\omega)$ and calculate $C_0$ from this. $\Phi(\omega)$ is
defined as the fourier transform of the velocity co-variance
function $R_L(\tau)$~\citep{tennekes}:
\begin{equation}
\Phi(\omega)=\frac{1}{2 \pi} \int_{-\infty}^{\infty} d\tau
\exp{(-\imath \omega \tau)} R_L(\tau).
\end{equation}
In the inertial range K41 predicts
\begin{equation}
\Phi_{ij}(\omega)=\beta \varepsilon \omega^{-2},
\end{equation}
with $C_0=\pi \beta$. In Figure \ref{fig:Ru} (a) we have plotted
$R_L(t)$ in the three directions. The radial components fall off
exponential with e-folding times $T_{exp}^x\sim 10.7 \tau_{\eta}$
and $T_{exp}^y\sim 9.4 \tau_{\eta}$ while the vertical
axisymmetric component $T_{exp}^z\sim 14 \tau_{\eta}$. Since
$R_L(\tau)$ is composed of eddies of all size in the flow, the
energy containing scales and hence the large scale inhomogeneities
strongly effects its form. The integration of $R_L(t)/\sigma^2$
gives the Lagrangian integral time scale $T_L$. We find values of
$T_L\sim T_{exp}$.
The velocity spectrum $\Phi(\omega)$ is shown in (b). For small
frequencies $\omega$ the spectrum for all three components are
white. This corresponds to uncorrelated velocities for long time
lags on a track. For higher frequencies all three spectra turn red
with slope of $\sim-2$ in agreement with the Kolmogorov
prediction. Due to a relative low sampling rate
($dt=0.021\mathrm{s}$) the Nyquist frequency prevent us from
studying frequencies higher than $\omega=23.8\mathrm{s^{-1}}$.

\begin{figure}[h]
\begin{center}
\includegraphics[width=\columnwidth]{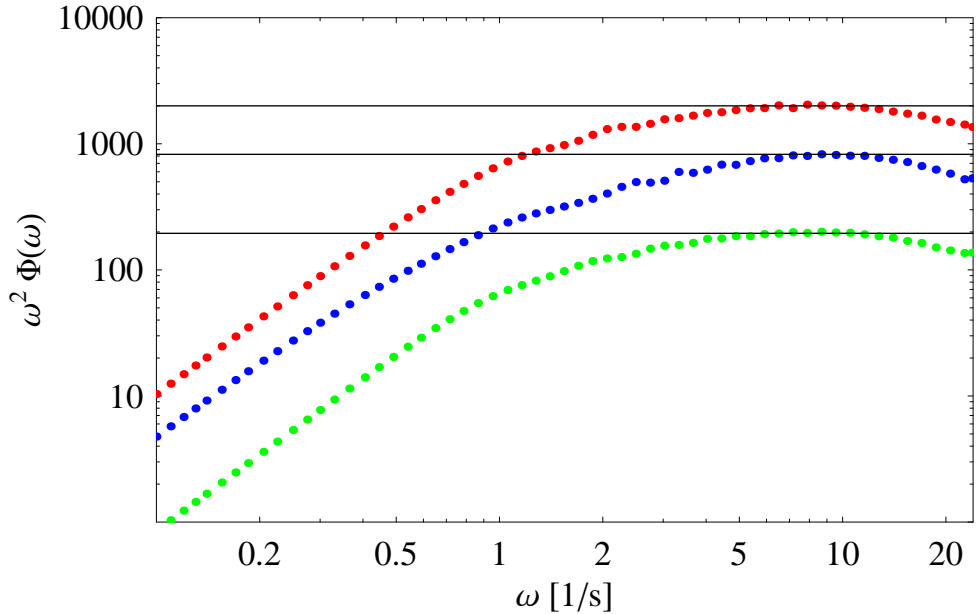}
\caption{{\footnotesize Compensated velocity spectrum $\omega^2
\Phi(\omega)$. Color code as in Figure~\ref{fig:C0}. The curves
have been shifted vertically for clarity. One can therefore not
determine the magnitude of $\omega^2 \Phi(\omega)$ from the
different curves. The horizontal lines are the levels from which
$C_0$ is calculated. }} \label{fig:compensatedSpectrum}
\end{center}
\end{figure}

\citet{lien:2002} studied the scaling properties in a simple
Lorentzian model spectrum and found that with a finite Reynolds
number it is easier to obtain inertial range scaling from the
spectrum than from the structure function $S_L^2(\tau)$. We have
plotted the spectrum compensated with $\omega^2$ in
Figure~\ref{fig:compensatedSpectrum} in order to have a better
look at the existence of an inertial range. For all three
components a narrow inertial range is observed as a plateau. The
horizontal lines are used for estimating $C_0$. We find values
equal to $4.91\pm0.15$, $4.79\pm0.14$ and $4.07\pm0.12$ for the
three components respectively. These values are smaller and a bit
more isotropic than those calculated from the structure functions.
This is in contrast to the arguments by \citet{lien:2002} $C_0$
who state that they should be larger.

\section{Small-scale intermittency}
From the study of the lower moments we proceed to higher order
moments describing the most extreme events.

\begin{figure}[h]
\begin{center}
\includegraphics[width=\columnwidth]{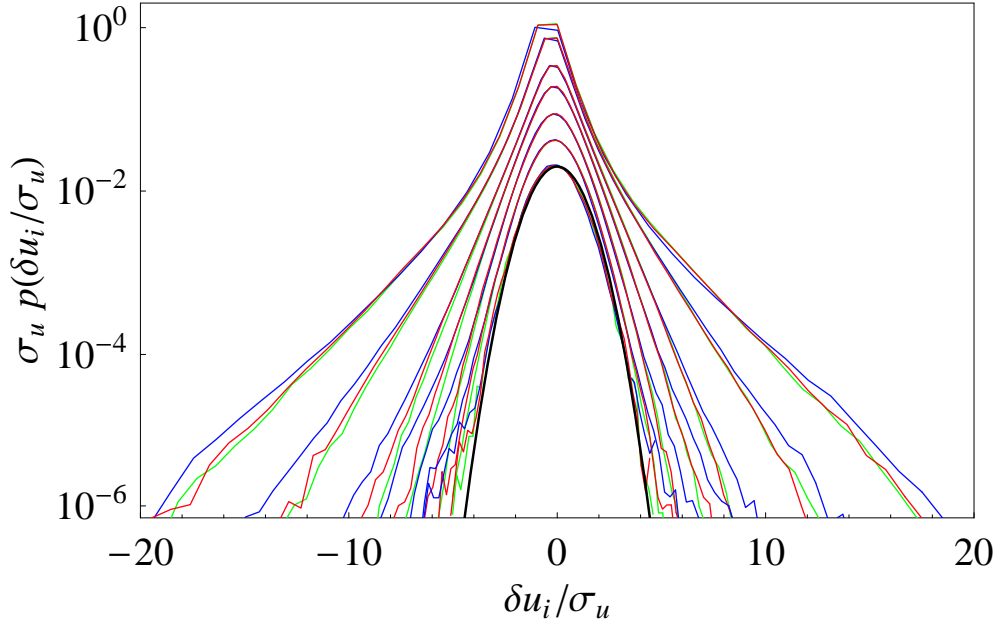}
\caption{{\footnotesize pdf of velocity increments $\delta
v(\tau)$ for times (decreasing downwards)
$\tau=1.0\tau_{\eta},1.7\tau_{\eta},3.6\tau_{\eta},6\tau_{\eta},12\tau_{\eta},24\tau_{\eta}$
and $48\tau_{\eta}$. The curves have been shifted vertically for
clarity. Color coding as in Figure~\ref{fig:C0}. The black curve
is a Gaussian.}} \label{fig:pdfDV}
\end{center}
\end{figure}

The pdfs of temporal velocity increments $\delta
v(\tau)=v(t+\tau)-v(t)$ are shown in Figure~\ref{fig:pdfDV} for
different time lags $\tau$. All three components are shown.
The three components show the same over all behavior: for large
time lags the distributions are Gaussian while they for small time
lags have fat tails. The curves corresponding to the smallest time
lags have a flat plateau at $\delta v\sim 0$. This is a binning
artifact and does therefore not represent any physical trend in
the data. The non-Gaussianity for small times becomes more clear
by looking at the flatness.
\begin{figure}[h]
\begin{center}
\includegraphics[width=\columnwidth]{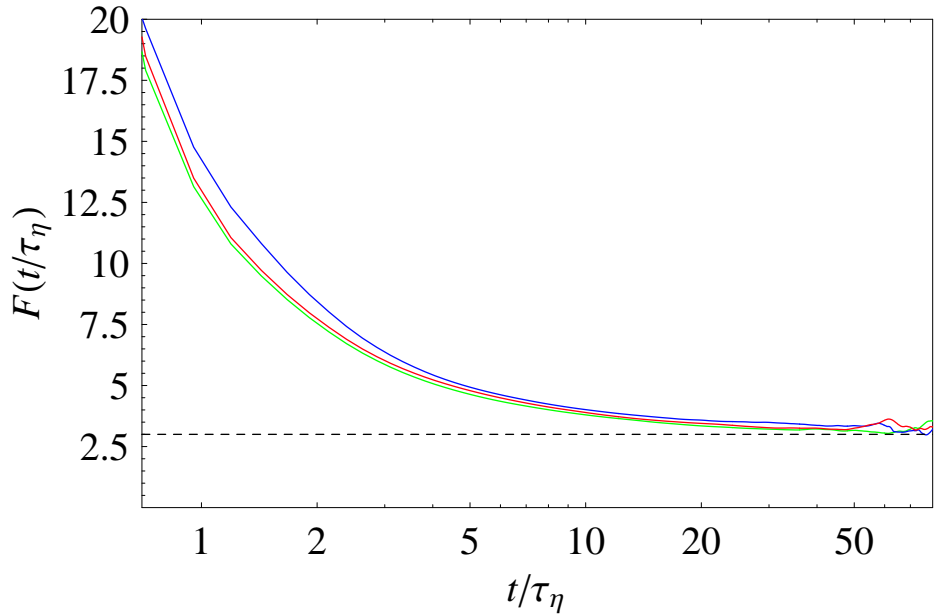}
\caption{{\footnotesize The flatness of $\delta v(\tau)$. The
color coding as in Figure~\ref{fig:C0}. The horizontal line $F=3$
is the Gaussian prediction.}} \label{fig:pdfM4}
\end{center}
\end{figure}
The flatness of the distributions is defined as
\begin{equation}
F(\tau)\equiv \frac{\langle \delta v^4(\tau) \rangle}{\langle
\delta v^2(\tau)\rangle^2}
\end{equation}
and is shown in Figure~\ref{fig:pdfM4}. $F$ is monotonically
decreasing for all three components and reaches a Gaussian level
at time lags: $\tau \sim 40\tau_{\eta}$, which is substantial
larger than $T_L$. We do not at present have any explanation for
this.

The results presented in Figure~\ref{fig:pdfDV} and
\ref{fig:pdfM4} are strong evidence of Lagrangian intermittency,
i.e. non-Gaussian behavior of the smallest temporal scales in the
flow. These results agree with observations by
\citet{mordant:2001} and Direct Numerical Simulations (DNS) by
\citet{biferale:2006}.

Our findings suggest that intermittency can be studied in flows
with a moderate Reynolds number of order $\mathcal{O}(100)$. The
only necessary condition seems to be the size of the ensemble: a
large number of particles is needed to observe rare events.

\subsection{Higher order structure functions and ESS}
\begin{figure}[h]
\begin{center}
\includegraphics[width=\columnwidth]{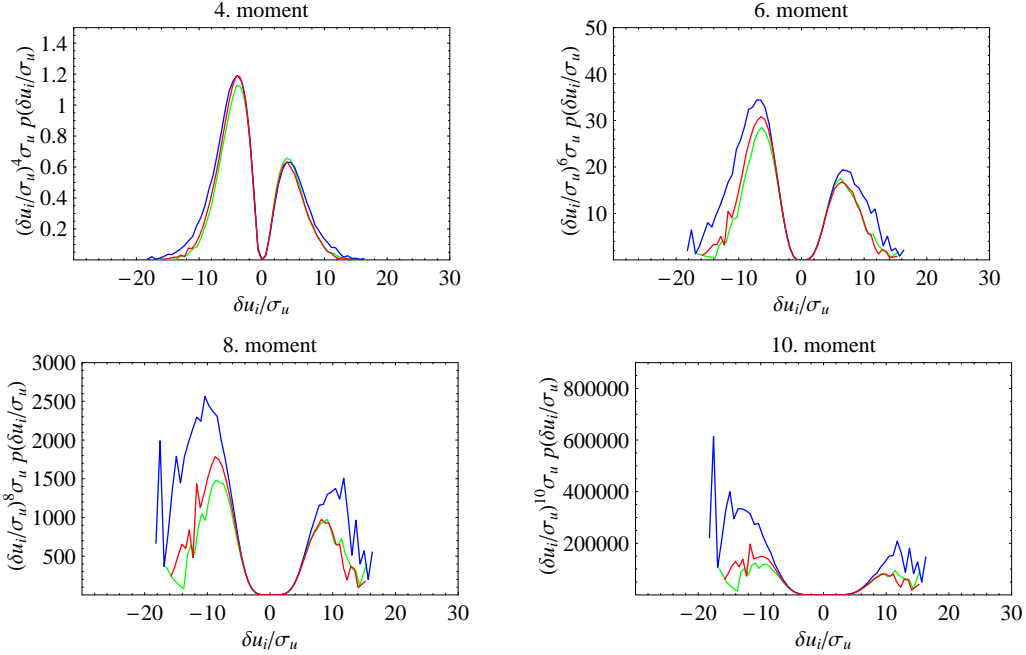}
\caption{{\footnotesize $\delta v^n p(\delta v)$ for $n=4,6,8,10$.
The time in all four plot is $\tau=2.1\tau_{\eta}$.}}
\label{fig:convergence}
\end{center}
\end{figure}

Before we look at the higher order moments we check for
convergence of these. In Figure~\ref{fig:convergence} we show
$\delta v^n(\tau) p(\delta v(\tau))$ for $n=4,6,8,10$. The time
lag in all four plot is $\tau=2.1\tau_{\eta}$.
For $n<8$ we observe convergence. For $n=8$ we start to get into
trouble, but it seems like we have captured most of the signal --
at least for the radial components (red and green curves). In an
incompressible flow $\langle \delta u(\tau) \rangle =0$: the
non-zero skewness observed in all the curves is therefore an
artifact of sampling errors and / or tracking of particles. This
is an issue which has to be resolved before more dramatic
conclusions can be made.

K41 similarity theory predicts in the inertial range for the $p$
order structure function:
\begin{equation}
S^p_L(\tau)\equiv\langle \delta v^p(\tau) \rangle \sim
\varepsilon^{p/2} \tau^{p/2}. \label{k41LagScaling}
\end{equation}
Intermittency can be defined as the departure from K41 similarity
scaling. This means that eqn.~\ref{k41LagScaling} can be replaced
by a more general form taking intermittency into account:
\begin{equation}
\langle \delta v^p(\tau)\rangle \sim \tau^{\zeta_p^L},
\label{intermittencyLagScaling}
\end{equation}
where $\zeta_p^L$ is commonly  named the Lagrangian anomalous
scaling exponent. Only recently it has been possible to measure
$\zeta_p^L$ and hence quantitatively describe the extreme dynamics
present in the fat tails of the distribution of $\delta v(\tau)$
for $\tau \rightarrow 0$
\citep{mordant:2001,mordant:2004b,biferale:2004,xu:2006,xu:2006b}.
The data presented here is therefore merely a verification of
already obtained results.

\begin{figure}[h]
\begin{center}
\includegraphics[width=\columnwidth]{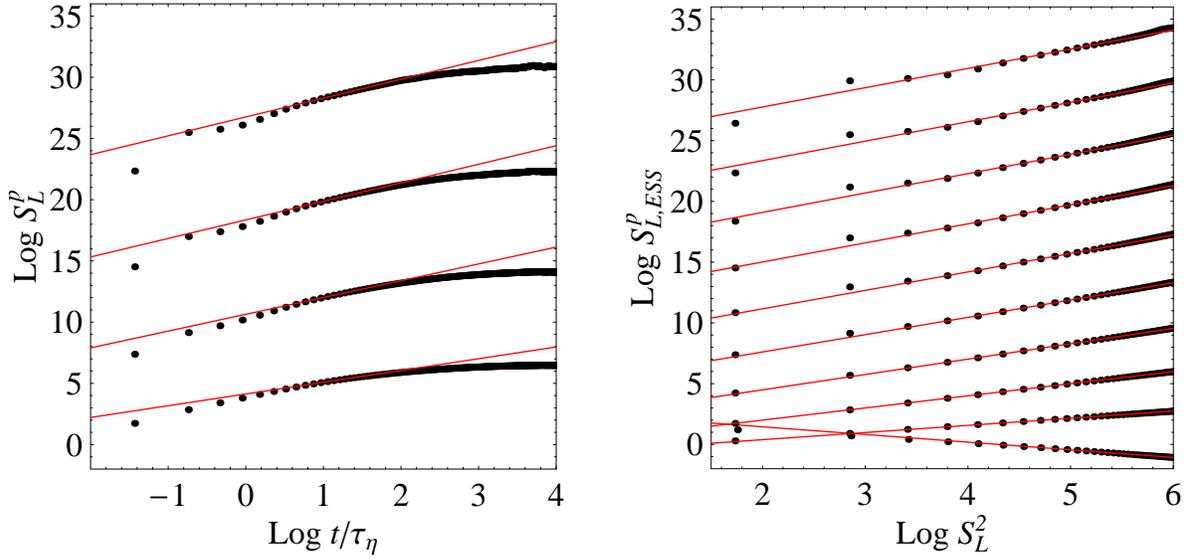}
\caption{{\footnotesize (a) $S_L^p(t)$ as a function of
$t/\tau_{\eta}$ for $p=2,4,6,8$ increasing upwards. (b) Extended
self-similarity: $S^p_{L,ESS}(t)$ as a function of $S_L^2(t)$ for
$p=-1,1,2,...,9$. In both panels data from the radial
$x$-component are used.}} \label{fig:scalingHO}
\end{center}
\end{figure}

In Figure \ref{fig:scalingHO} (a) structure functions $S_L^p(t)$
of order $p=2,4,5,6,8$ are shown as a function of $\tau$. Power
laws have been fitted to each function in the region
$2\tau_{\eta}\leq t \leq 4\tau_{\eta}$ corresponding to the maxima
of $S_L^p(t)/t/\varepsilon$. The fits are not convincing. First of
all, the inertial range is too narrow and we therefore can not
expect any universal scaling. Secondly, and less importantly, we
know that the small scales are affected by the large-scale
inhomogeneities.

\begin{figure}[h]
\begin{center}
\includegraphics[width=\columnwidth]{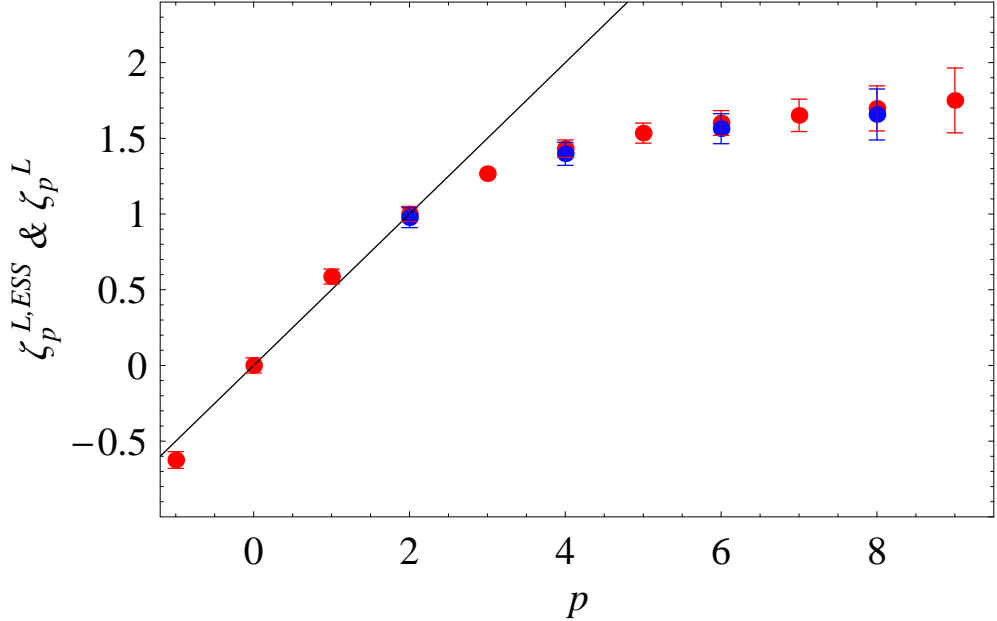}
\caption{{\footnotesize Lagrangian anomalous scaling exponents
$\zeta^L_p$ (blue curve) and the Extended self-similarity
anomalous scaling exponents $\zeta^{L,ESS}_p$ (red curve). The
straight line is the K41 prediction.}} \label{fig:exponents}
\end{center}
\end{figure}

A popular way of looking at scaling exponent is instead to measure
ratios of scaling exponents. This method is called Extended
self-similarity (ESS) and was introduced by \citet{benzi:1993}.
The method was introduced in the Eulerian frame but can be
transferred to the Lagrangian frame if we assume that
$\zeta_2^L=1$ following K41 similarity theory. The crucial step is
to treat all velocity increments as positive. This affects the
odd-numbered structure functions. We therefore define
\begin{equation}
S_{L,ESS}^p(\tau)\equiv\langle | \delta v(\tau) |^p \rangle \sim
\langle \delta v^2(\tau) \rangle^{\zeta^{L,ESS}_p}.
\end{equation} In Figure \ref{fig:scalingHO} (b) $S^p_{L,ESS}(t)$
is shown as a function of $S_L^2(t)$. The scaling is now much
better, which explains the wide popularity of the method. The
different scaling exponents are printed in Table~\ref{table:zeta}
and plotted in Figure~\ref{fig:exponents} for the radial
components. The error bars represent small deviations between the
two radial components as well as an error estimated from fitting
the straight lines in Figure~\ref{fig:scalingHO}. The errors
increase with $p$ and are significantly larger for $\zeta^L_p$
compared to the ESS approach $\zeta^{L,ESS}_p$.

\begin{table}[t]
\begin{tabular}{|l|c|c|c|c|c|}
\hline
$p$ & $-1$ & $1$ &$2$ &$3$ &$4$  \\
  \hline\hline
 $\zeta^L_p$ & $-$ &$-$ & $0.98\pm0.07$ & $-$ & $1.40\pm0.08$ \\
  \hline
  $\zeta^{L,ESS}_p$ & $-0.62\pm0.07$ & $0.59\pm0.02$ & $1$ & $1.27\pm0.03$ & $1.43\pm0.06$  \\
  \hline
\end{tabular}
\vspace*{0cm}
\begin{tabular}{|l|c|c|c|c|c|}
\hline
$p$ &$5$  &$6$&$7$&$8$&$9$ \\
  \hline\hline
 $\zeta^L_p$ & $-$ & $1.56\pm0.10$ & $-$ & $1.66\pm0.17$ & $-$  \\
  \hline
  $\zeta^{L,ESS}_p$ & $1.53\pm0.05$ & $1.60\pm0.06$ & $1.65\pm0.09$ & $1.70\pm0.13$ & $1.75\pm0.19$ \\
  \hline
\end{tabular}
\caption{{\footnotesize Lagrangian anomalous scaling exponents}}
\label{table:zeta}
\end{table}

Some remarks about ESS should be made at this point. In the
original paper \citet{benzi:1993} argued, based on experimental
evidence of $|\langle \delta_r u^3(r) \rangle| \sim \langle
|\delta_r u(r)|^3 \rangle,$ that $\langle |\delta_r u(r)|^p
\rangle = B_p \langle |\delta_r u(r)|^3
\rangle^{\zeta^{E,ESS}_p}$. As also emphasized in the paper this
is not a rigorous result which can be deduced from the
Navier-Stokes equation. By plotting absolute (defined by positive
velocity increments) structure functions vs. the third order
structure function (Eulerian frame) or the second order structure
function (Lagrangian frame), an extended scaling range can be
observed because undulations in the structure functions are
correlated and hence disappear when plotted against each other.
ESS is widely used and gives seemingly universal scaling exponents
for flows in a wide range of Reynolds numbers. As pointed out by
\citet{arneodo:1996} no consensus besides the observed facts
exists about the interpretation or even significance of ESS.
Whether the observed scaling in ESS is the signature of hidden
physical laws is speculated. In the Lagrangian frame an additional
problem arise. As already mentioned K41 predicts linear dependence
of time scale for the Lagrangian second order structure function
and hence $\zeta_2^L=1$. This is motivated by the scaling in the
Eulerian frame and specifically from the \textit{four-fifth} law.
A similar \textit{exact} result does not exist for the Lagrangian
structure functions. So all in all, one could state that
\textit{it is a wonder that it works!}

The values in Table~\ref{table:zeta} are in excellent agreement
with results obtained by \citet{xu:2006b} and
\citet{mordant:2004b}. The values by \citet{biferale:2004} are
somehow higher and was by \citet{xu:2006} explained as a different
choice of inertial range.

\subsection{The multifractal framework}
The multifractal model of turbulence was introduced by
\citet{parisi:1985} in the Eulerian frame after an early attempt
by \citet{Mandelbrot:1975} who used multifractal measures to
characterize the spatial structure of dissipation in turbulence.

The multifractal model is phenomenological and has been able to
successfully predict the corrections to K41 similarity scaling for
high order moments of spatial velocity increments
\citep{meneveau:1990,frisch,sreenivasan:1997}.

\citet{borgas:1993} discusses multifractals in the Lagrangian
frame and introduces a bridge to the Eulerian framework. The
literature is, however, not very rich on work on Lagrangian
multifractals, which could have to do with the difficulties in
obtaining reliable Lagrangian data set more than a animosity
against the multifractal model. Work by
~\citet{biferale:2004,biferale:2005b,chevillard:2003,mordant:2002,mordant:2004b,xu:2006b}
have, however, shed light on the issue of multifractals in the
Lagrangian frame.

In the Lagrangian multifractal model the flow is assumed to
possess a range of scaling exponents $h_{min},...,h_{max}$ with a
certain probability so that
\begin{equation}
\delta v(\tau) \sim \tau^h.
\end{equation}
For each scaling exponent $h$ there is a fractal set with a
$h$-dependent dimension $D^L(h)$. The embedding dimension is one
($\tau\in\mathcal{R}$) and hence $D^L(h)\leq1$ for all $h$. The
probability $P^L_h(\tau)$ of having an exponent $h$ at time $\tau$
is therefore proportional to $1-D^L(h)$. From a steepest descent
argument one can calculate a relation between the anomalous
scaling exponents $\zeta^L_p$ and the fractal dimension $D^L(h)$
given by~\citep{frisch}:
\begin{equation}
\zeta^L_p=\underset{h}{\inf}[ph+1-D^L(h)] \label{eq:L1}.
\end{equation}
If $D^L(h)$ is \textit{concave} a Legendre transformation gives
\begin{equation}
D^L(h)=\underset{p}{\inf}[ph+1-\zeta^L_p] \label{eq:L2}.
\end{equation}

\begin{figure}[h]
\begin{center}
\includegraphics[width=\columnwidth]{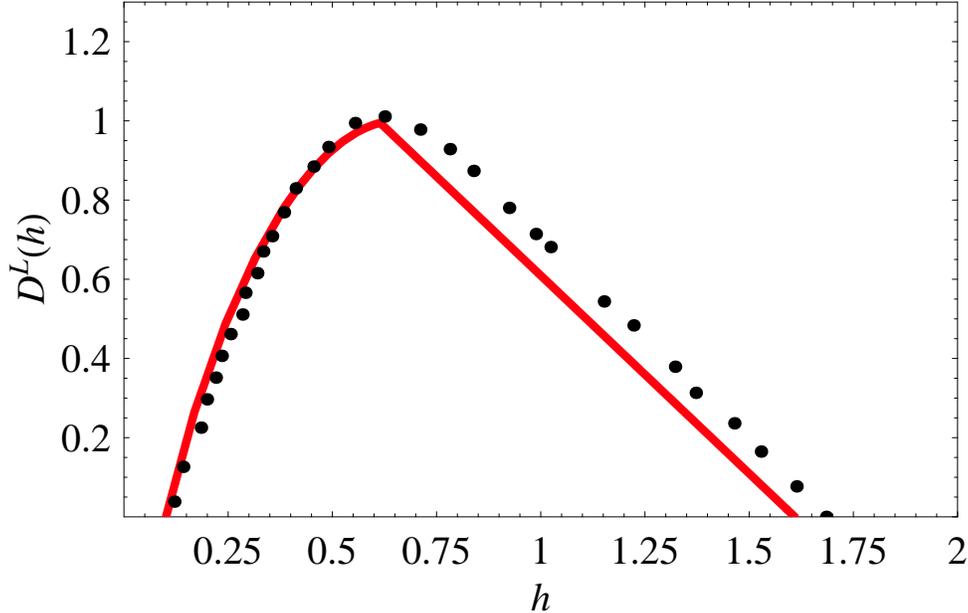}
\caption{{\footnotesize The Lagrangian multifractal dimension
spectrum $D^L(h)$. The black dots are the result by
\citet{xu:2006}.}} \label{fig:Dh}
\end{center}
\end{figure}

In Figure \ref{fig:Dh} we have plotted $D^L(h)$ obtained through
eqn.~\ref{eq:L2}. First we calculated $\zeta^L_p$ for both integer
and non-integer values of $p$ between $p=-1$ and $p=9$. The result
is the red curve in the Figure~\ref{fig:Dh}.

The black dots are the result by \citet{xu:2006} who in a PTV
experiment of Reynolds number $Re_{\lambda}=200,690$ and $815$
measured $D^L(h)$ both trough $P^L_h(\tau)$ which they manage to
measure directly and through eqn.~\ref{eq:L2} as we have done
here. They arrived at the same $D^L(h)$ from both calculations
putting confidence in the multifractal model for Lagrangian
velocity increments. The agreement between their data and ours is
very good. Only for $h>0.6$ in the linear portion of $D^L(h)$ do
we observe a discrepancy.

This linear portion of $D^L(h)$ was by \citet{xu:2006} explained
in the following way: because the domain of $h$ is finite
eqn.~\ref{eq:L1} will become a linear function after some $p'$.
This linear behavior is also observed in
Figure~\ref{fig:exponents} for large $p$s. The transition point
$h'$ happens where $p'$ minimizes the right hand side of
eqn.~\ref{eq:L1}. For $p>p'$ we therefore have that
$\zeta^L_p=h_{min} p+1$. Since only moments of the structure
functions of order larger than $-1$ converge we have $p'=-1$ and
the linear part of the curve is $D^L(h)=-h+1-\zeta_{-1}$.
\citet{xu:2006} successfully corrected the models by
\citet{biferale:2004} (from a theoretical prediction by
\citet{she:1994}) and \citet{chevillard:2003} and found a
remarkable match. The discrepancy in Figure \ref{fig:Dh} therefore
stems from different estimates of $\zeta^L_{-1}$ and the
uncertainty in measuring it.

\begin{figure}[h]
\begin{center}
\includegraphics[width=\columnwidth]{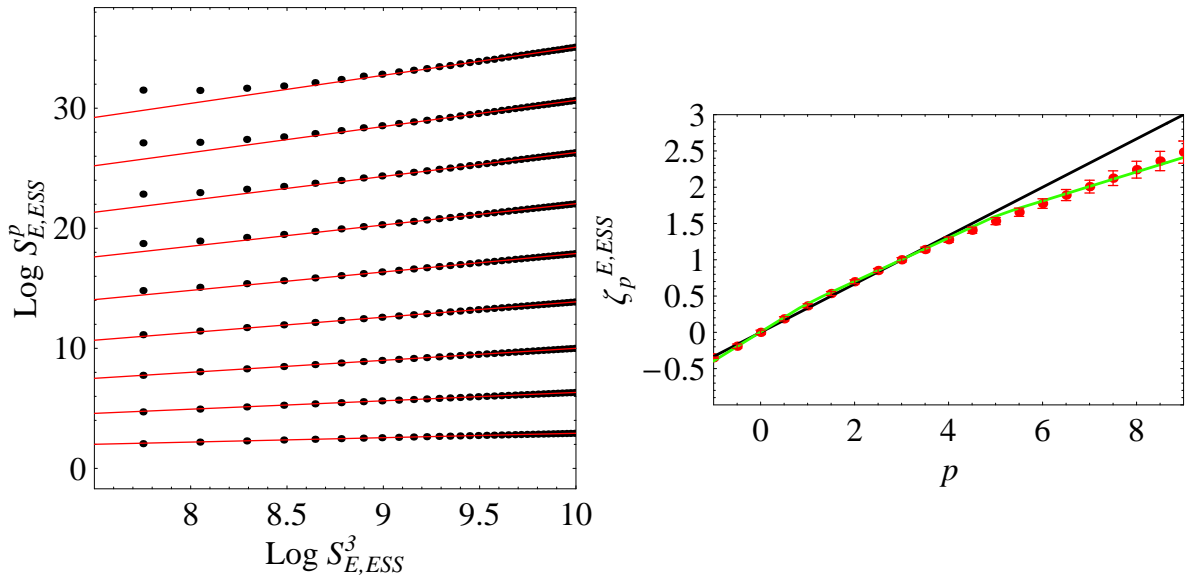}
\caption{{\footnotesize (a) ESS Eulerian structure functions
$S^p_{E,ESS}(r)$ of order $p$ as a function of $S^3_E(r)$.
$p=1,..,9$ increasing upwards. (b) Anomalous scaling exponent
determined from (a) (red dots). The straight line is the K41
prediction and the green curve is the theoretical model by
\citet{she:1994}.}} \label{fig:DhEUL}
\end{center}
\end{figure}

\citet{chevillard:2003} came up with a formula for the connection
between $D^L(h)$ and its Eulerian counterpart $D^E(h)$. The
formula is
\begin{equation}
D^L(h)=-h+(1+h)\left(D^E\left(\frac{h}{1+h}\right)-2\right).
\label{eq:transformation}
\end{equation}
From our database we have calculated the Eulerian anomalous
scaling exponents from ESS structure functions
\begin{equation}
S^p_{E,ESS}(r)\equiv\langle |\delta_r u(r)|^p \rangle \sim \langle
|\delta_r u(r)|^3 \rangle^{\zeta^{E,ESS}_p}.
\end{equation}
Results are shown in Figure~\ref{fig:DhEUL}. In (a)
$S^p_{E,ESS}(r)$ are plotted from $p=1,...9$. For all orders ESS
seems to work fine. In (b) the anomalous scaling exponents
$\zeta^{E,ESS}_p$ are shown.

The values are in perfect agreement with the theoretical model by
\citet{she:1994}. More interesting is the departure from the K41
prediction which is smaller than in the Lagrangian frame. This is
interpret as Lagrangian statistics being more intermittent.

Just like in the Lagrangian frame there is a Legendre
transformation between $\zeta^E_p$ and $D^E(h)$:
\begin{equation}
D^E(h)=\underset{p}{\inf}[ph+3-\zeta^E_p] \label{eq:E1}
\end{equation}
The only difference from eqn.~\ref{eq:L2} is the embedding
dimension which in the Eulerian frame is three $(\mathbf{r}\in
\mathcal{R}^3).$

\begin{figure}[h]
\begin{center}
\includegraphics[width=\columnwidth]{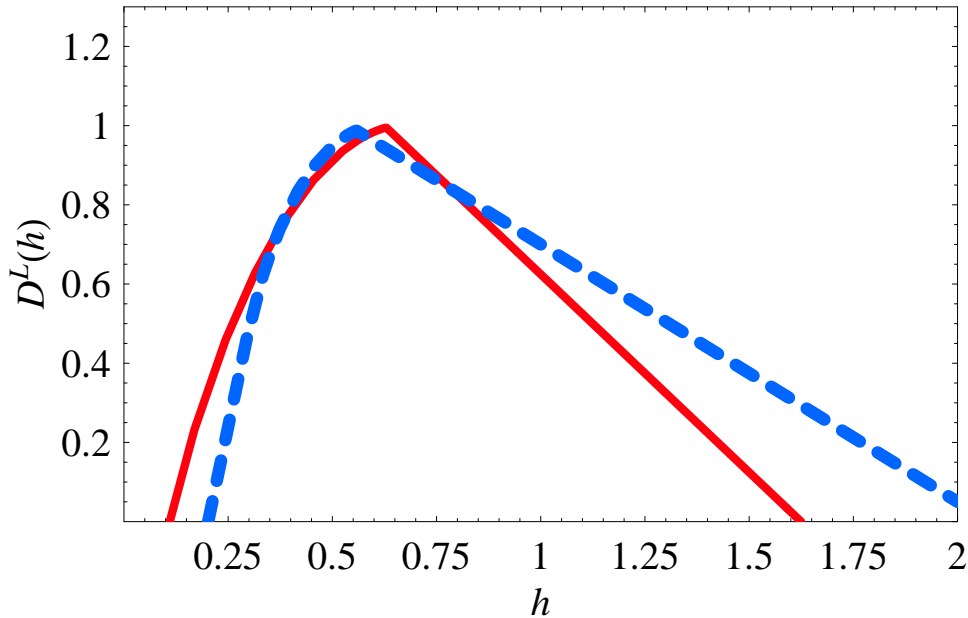}
\caption{{\footnotesize The Lagrangian multifractal dimension
spectrum $D(h)$. The red curve is calculated from Lagrangian ESS
structure functions while the blue is obtained though Eulerian
structure functions and \ref{eq:transformation}.}}
\label{fig:DhCOMP}
\end{center}
\end{figure}

From eqn.~\ref{eq:E1} and ~\ref{eq:transformation} we can find
$D^L(h)$ from the Eulerian anomalous scaling exponent presented in
Figure \ref{fig:DhEUL} (b). The comparison is plotted in Figure
\ref{fig:DhCOMP}. Again we observe a discrepancy in the linear
part of $D^L_h$. Whether it comes from the determination of the
anomalous scaling exponents from ESS or that there is a flaw in
eqn.~\ref{eq:transformation} we can not say at the moment. A
direct measurement of the probability $P_h$ in both the Eulerian
and Lagrangian frame might give more insight into the connection
between the two frames.

The physical interpretation of the multifractal model is not that
easy. In K41 similarity scaling only one scaling exponent is
possible, namely $h=1/3$ and thus $\zeta^E_p=p/3$. This is
motivated from the fact the Navier-Stokes (N-S) equation is only
invariant under one scaling group. This group is characterized by
an exponent $\bar{h}$ obtained by scaling the N-S with the
following transformation (time,position,velocity):
$t,\mathbf{r},\mathbf{u} \mapsto
\lambda^{1-\bar{h}}t,\lambda\mathbf{r},\lambda^{\bar{h}}\mathbf{u}$
for $\lambda \in \mathcal{R}_+$. The solution is $\bar{h}=-1$. In
the limit of infinite Reynolds number the viscosity term in the
N-S equation becomes negligible and we find that the N-S equation.
is now invariant to infinitely many exponents $\bar{h}$. This is
one of the motivations for the multifractal model. It is, however,
not a justification. Another important aspect of the model is the
fact that when an eddy breaks up into smaller eddies in the
Richardson picture the smaller eddies do not cover the same amount
of space. Instead they cover only a fraction equal to $3-D^E(h)$.
We thus have regions in the flow with large activity and regions
with almost calm waters. In the Lagrangian frame this would mean
that the individual fluid particles are \textit{not free} to move
around in all directions. For example as reported by
\cite{laporta:2001} and \cite{biferale:2005b} are particles often
trapped by intense vortices. The large accelerations and velocity
increments of these events are therefore of dimension lower than
$3$ in the Eulerian frame and lower than $1$ in the temporal. This
spiral motion of fluid particle around a fluid filament is also
the fluid mechanical picture of intermittent events in the model
by \citet{she:1994}: by entraining surrounding fluid kinetic
energy fluctuations are effectively dissipated along the axis of
the filament.

As emphasized by \citet{borgas:1993} the multifractal model does,
however, not imply that the trajectories of fluid particles are
fractal trajectories themselves.

\section{Conclusions}
We have measured Lagrangian one-particle statistics and looked at
small-scale behavior. The finiteness of the measuring volume can
be used to calculate the mean kinetic energy dissipation
$\varepsilon$ in the flow without any further assumptions. The
small scales do seem to be affected by the large-scale
inhomogeneities present in our flow. We do not observe a
significant inertial range but by Extended Self-Similarity we are
able to extract a quantitative measure of the structure functions
of high order. From these we calculate the Lagrangian anomalous
scaling exponents and find excellent agreement with already
published results.

Via the multifractal model we have calculated the Lagrangian
multifractal dimension spectrum. The spectrum is similar to the
one published by \citet{xu:2006b} even though our Reynolds number
is significantly lower and our mean flow is different.

Most importantly we have shown that a high Reynolds number is not
necessary to obtain results in the Lagrangian frame. All
experiments and DNS do show the same qualitative features and no
clear Lagrangian inertial range has been observed. Whether it is
because current experiments are performed with too low Reynolds
number or it simply do not exist future experiments will tell.
\acknowledgments The author is grateful to Beat L\"{u}thi,
{S{\o}ren Ott} and Jakob Mann.
\bibliographystyle{plainnat}

\end{document}